\documentstyle[12pt]{article}
\begin{document}
\begin{titlepage}
\begin{flushright}
IC/2001/21\\
hep-th/0104011
\end{flushright}
\vspace{10 mm}

\begin{center}
{\Large A Note on Thermodynamics and Holography of\\ 
Moving Giant Gravitons}

\vspace{5mm}

\end{center}

\vspace{5 mm}

\begin{center}
{\large Donam Youm\footnote{E-mail: youmd@ictp.trieste.it}}

\vspace{3mm}

ICTP, Strada Costiera 11, 34014 Trieste, Italy

\end{center}

\vspace{1cm}

\begin{center}
{\large Abstract}
\end{center}

\noindent

In our previous work (Phys. Rev. D63, 085010, hep-th/0011290), we showed 
that the brane universe on the giant graviton moving in the near-horizon 
background of the dilatonic D$(6-p)$-brane is described by the mirage 
cosmology.  We study thermodynamic properties of the moving giant graviton 
by applying thermodynamics of cosmology and the recently proposed holographic 
principles of cosmology.  We find that the Fischler-Susskind holographic 
bound is satisfied by the closed brane universe on the moving giant graviton 
with $p>3$.  The Bekenstein and the Hubble entropy bounds and the recently 
proposed Verlinde's holographic principle applied to the brane universe on 
the giant graviton are also studied.

\vspace{1cm}
\begin{flushleft}
April, 2001
\end{flushleft}
\end{titlepage}
\newpage

It is argued in Ref. \cite{suss} that gravitons moving in the $S^n$ portion 
of the AdS$\times S^n$ spacetime blow up into spherical $(n-2)$-brane of 
increasing size with increasing angular momentum through the Myers effect 
\cite{myr}.  Later, it is further argued that gravitons can expand into a 
sphere also in the AdS portion \cite{ads1,ads2} of the AdS$\times S^n$ 
background and even in the background spacetime other than the AdS$\times S^n$ 
spacetime provided certain conditions on the bulk fields are satisfied 
\cite{dtv}.  Their argument is based on the fact that the spherical probe 
$p$-brane with the stable radius and nonzero angular momentum wrapping the 
$S^p$ portion of a certain  background spacetime has the same energy as that 
of a graviton with the same angular momentum.  
The giant graviton moving in the near horizon background of the dilatonic 
D-brane is shown \cite{youm} to be described by the expanding closed universe 
in mirage cosmology
\footnote{The term `mirage' was coined to signify that the expansion of the 
brane universe is not due to energy density and pressure of real matter 
on the brane, but due to something else, namely, the geodesic motion of 
the universe brane in the curved background of other brane(s).} 
\cite{cr,cpr,kr,kk}.  Recently, there has been growing 
interest in various holographic bounds \cite{sus,esl,ven,bar,kal,bou,ver1} 
in cosmology.  It is the purpose of this note to apply such conjectured 
holographic bounds to the mirage cosmology on the moving giant graviton to 
infer thermodynamic properties of the moving giant graviton and its 
holographic dual theory.  

We begin by briefly summarizing the result of our previous work.  The 
nontrivial fields for the near-horizon background of the source 
D$(6-p)$-brane, magnetically charged under the RR $(p+1)$-form potential 
$A^{(p+1)}$, have the forms:
\begin{eqnarray}
ds^2&=&-g_{tt}dt^2+\sum^{6-p}_{i=1}g_{ii}dx^2_i+g_{rr}dr^2
+h(r)r^2d\Omega^2_{p+2},
\cr
e^{\Phi}&=&\left(\textstyle{L_p\over r}\right)^{{(p-3)(p+1)}\over 4},
\ \ \ \ \ \ 
A^{p+1}_{\phi\theta_1...\theta_p}=L^{p+1}_p\rho^{p+1}
\epsilon_{\theta_1...\theta_p},
\cr
g_{tt}&=&\left(\textstyle{r\over L_p}\right)^{{p+1}\over 2},\ \ \ 
g_{ii}=\left(\textstyle{r\over L_p}\right)^{{p+1}\over 2},\ \ \ 
g_{rr}=\left(\textstyle{L_p\over r}\right)^{{p+1}\over 2},\ \ \ 
h(r)=\left(\textstyle{L_p\over r}\right)^{{p+1}\over 2},
\label{nrdbrnback}
\end{eqnarray}
where $\epsilon_{\theta_1...\theta_p}$ is the volume form of a unit 
$S^p$ and the metric $d\Omega^2_{p+2}$ is parametrized as
\begin{equation}
d\Omega^2_{p+2}={1\over{1-\rho^2}}d\rho^2+(1-\rho^2)d\phi^2+\rho^2
d\Omega^2_p.
\label{sdp2}
\end{equation}
When the spherical probe $p$-brane with the tension $T_p$ and the angular 
momentum ${\cal P}_{\phi}$ has the stable (coordinate) radius $\rho_0=
({\cal P}_{\phi}/N)^{1/(p-1)}$, its energy ${\cal E}$ is the same as that 
of a graviton with the same angular momentum ${\cal P}_{\phi}$ and thereby 
the probe brane can be identified as a graviton blown up into a sphere, i.e., 
the giant graviton.  Here, $N\equiv T_pV_pL^{p+1}_p$ is the number of 
D$(6-p)$-branes, where $V_p\equiv 2\pi^{{p+1}\over 2}/\Gamma\left({{p+1}\over 
2}\right)$ is the volume of the unit $S^p$.  The induced metric on the 
spherical probe $p$-brane can be put into the following standard comoving 
frame form for the Robertson-Walker metric for the expanding closed universe:
\begin{equation}
d\tilde{s}^2=-d\eta^2+a^2(\eta)d\Omega^2_p
=-d\eta^2+a^2(\eta)\left[d\theta^2+\sin^2\theta d\Omega^2_{p-1}\right],
\label{indmet}
\end{equation}
where $\eta$ is the cosmic time, $\theta$ is the azimuthal angle, 
$d\Omega^2_{p-1}$ is the line element on the unit polar $(p-1)$-sphere, and 
$a(\eta)\equiv\sqrt{h(r)}r\rho_0=L^{(p+1)/4}_p\rho_0r^{(3-p)/4}$ is the 
cosmic scale factor.  Note, for the dilatonic D-brane case (i.e., the $p
\neq 3$ case), the volume of the spherical probe $p$-brane or the giant 
graviton varies while the giant graviton moves along the $r$-direction of 
the background spacetime (\ref{nrdbrnback}), although its coordinate radius 
$\rho_0$ remains constant.  When $p<3$ [$p>3$], the volume ($\sim a^p$) of 
the giant graviton increases [decreases] while the giant graviton moves away 
from the source D$(p-6)$-brane.  The time evolution of $a$, while the giant 
graviton moves along the $r$-direction, is described by the following 
Friedmann-like equations:
\begin{equation}
\left({\dot{a}\over a}\right)^2={{(p-3)^2}\over {16}}\left[\bar{\cal E}^2
\bar{\cal P}^{-2{{p-1}\over{p-3}}}_{\phi}a^{2{{p+1}\over{p-3}}}-{1\over a^2}
\right],
\label{frdeq1}
\end{equation}
\begin{equation}
{\ddot{a}\over a}={{(p-1)(p-3)}\over 8}\bar{\cal E}^2\bar{\cal P}^{-2{{p-1}
\over{p-3}}}_{\phi}a^{2{{p+1}\over{p-3}}},
\label{frdeq2}
\end{equation}
where $\bar{\cal E}\equiv {\cal E}/N$, $\bar{\cal P}_{\phi}\equiv 
{\cal P}_{\phi}/N$ and the overdots stand for derivatives w.r.t.  $\eta$.  
From Eq. (\ref{frdeq2}), we see that $\ddot{a}<0$ [$\ddot{a}>0$] for $p=2$ 
[$p>3$], for any values of $a$.  Since $a\sim r^{(3-p)/4}$, 
this implies that the giant graviton is always attracted to the dilatonic 
($p\neq 3$) source D$(6-p)$-brane.  For $p>3$ the volume of the giant 
graviton approaches infinity, whereas for $p=2$ its volume approaches zero, 
as it approaches the source D$(6-p)$-brane.

Spherical D-branes can be regarded as bound states of D0-branes.   
(See Refs. \cite{ars1,ars2,hns} for the spherical D2-brane case.)  
So, spherical D-branes, regarded as gas of D0-branes, are thermodynamic 
system described by thermodynamic quantities such as entropy, pressure and 
temperature.  Also, thermodynamics of the giant graviton can be regarded 
as being originated microscopically from the vibrating excitations arising 
from motion of the giant graviton in spacetime \cite{djm}.  Since the 
motion of the giant graviton is described by the Friedmann equations 
(\ref{frdeq1},\ref{frdeq2}) for an expanding closed universe, we would 
expect that thermodynamics on the giant graviton moving along the 
$r$-direction of the source brane emulates thermodynamics of the expanding 
closed universe.  

First of all, the energy density and the pressure on the moving giant 
graviton are expected to be given by the effective energy density and 
pressure determined from the effective Friedmann equations 
(\ref{frdeq1},\ref{frdeq2}).  The Friedmann equations for an expanding 
$(p+1)$-dimensional closed universe are generally expressed in terms of 
the energy density $\varrho$ and the pressure $\wp$ of the perfect fluid 
matter in the universe as
\begin{equation}
\left({\dot{a}\over a}\right)^2={{16\pi G}\over{p(p-1)}}\varrho-{1\over a^2},
\label{frd1}
\end{equation}
\begin{equation}
{\ddot{a}\over a}=-{{8\pi G}\over{p(p-1)}}\left[(p-2)\varrho+p\wp\right],
\label{frd2}
\end{equation}
where $G$ is the $(p+1)$-dimensional Newton's constant.  
So, the energy density and the pressure on the moving giant graviton are 
inferred to be given by
\begin{equation}
\varrho={{p(p-1)}\over{256\pi G}}\left[(p-3)^2\bar{\cal E}^2
\bar{\cal P}^{-2{{p-1}\over{p-3}}}_{\phi}a^{2{{p+1}\over{p-3}}}-(p+1)(p-7)
{1\over a^2}\right],
\label{efden}
\end{equation}
\begin{equation}
\wp=-{{p-1}\over{256\pi G}}\left[(p-3)(p^2-p+2)\bar{\cal E}^2
\bar{\cal P}^{-2{{p-1}\over{p-3}}}_{\phi}a^{2{{p+1}\over{p-3}}}-
(p-2)(p-7)(p+1){1\over a^2}\right].
\label{efprs}
\end{equation}
From this we see that the gas of D0-branes or the giant graviton is made up 
of noninteracting two species of perfect fluids satisfying the equations of 
state $\wp_i=w_i\varrho_i$ with $w_1=-{{p^2-p+2}\over{p(p-3)}}$ and 
$w_2=-{{p-2}\over p}$.  For the $p=2$ case, the gas of D0-branes is made 
up of perfect fluids with $w_1=2$ and $w_2=0$, namely the acausal fluid and 
the (pressureless) dust.  For the $p>3$ case, $w_1<0$ and $w_2<0$, so the 
pressure on the giant graviton is always negative, implying instability of 
the gas of D0-branes.  As a result, the brane universe on the giant graviton  
always accelerates ($\ddot{a}>0$), i.e., inflates, when $p>3$.  (Cf. It was 
previously shown \cite{tur,barr,gsb} that a gas of strings also has 
negative pressure, satisfying the equation of state $\wp=-\varrho/(D-1)$.)  

From the Friedmann equations (\ref{frd1},\ref{frd2}), we obtain the 
following conservation equation:
\begin{equation}
\dot{\varrho}+p(\varrho+\wp){\dot{a}\over a}=0.
\label{cnsveq}
\end{equation}
The second law of thermodynamics, applied to a comoving volume element of 
unit coordinate volume and physical volume $V=a^p$, implies the following 
entropy change for the giant graviton:
\begin{equation}
TdS=d(\varrho a^p)+\wp d(a^p).
\label{entchng}
\end{equation}
From Eqs. (\ref{cnsveq},\ref{entchng}), we see that the entropy of the gas 
of D0-branes per comoving volume stays constant in time, i.e., $dS/d\eta=0$.  
By substituting the following thermodynamic relation, obtained from the 
integrability condition $\partial^2S/(\partial T\partial V)=\partial^2S/
(\partial V\partial T)$:
\begin{equation}
{{d\wp}\over{dT}}={{\wp+\varrho}\over T},
\label{thmrel}
\end{equation}
into Eq. (\ref{entchng}), we obtain the following expression for the entropy 
of the gas of D0-branes per comoving volume:
\begin{equation}
S={{a^p}\over T}(\wp+\varrho)+S_0,
\label{entexp}
\end{equation}
where $S_0$ is an integration constant, which will be shown to be related 
to a Casimir entropy or energy.  So, substituting Eqs. 
(\ref{efden},\ref{efprs}) into Eq. (\ref{entexp}), we obtain the temperature 
of the giant graviton as a function of the cosmic scale factor $a$:
\begin{equation}
T={{p^2-1}\over{128\pi G(S_0-S)}}\left[(p-3)\bar{\cal E}^2
\bar{\cal P}^{-2{{p-1}\over{p-3}}}_{\phi}a^{{p^2-p+2}\over{p-3}}
+(p-7)a^{p-2}\right].
\label{tmpofd0}
\end{equation}
For $p=2$, $T$ scales as $a^{-4}$ for small $a$ and asymptotes to a constant 
as $a\to\infty$.  For $p>3$, $T$ scales as $a^{p-2}$ for small $a$, whereas 
it scales as $a^{(p^2-p+2)/(p-3)}$ for large $a$.  

The holographic principle of cosmology generally states \cite{sus} that the 
total entropy $S_{tot}=S{\cal V}$ inside a spherical region of coordinate 
volume ${\cal V}$ should not exceed the area $A$ of the bounding surface in 
Planck units, i.e., $S_{tot}/A<1$.  Note, we showed in the above that the 
homogeneous comoving entropy density $S$ is constant in $\eta$.  Fischler 
and Susskind (FS) \cite{sus} proposed that the holographic bound applies 
only to the part of the entropy contained within the particle horizon $L_H$, 
which is the product of $a$ and the comoving distance $r_H$ to the horizon.  
In the case of the closed universe, the comoving distance is given by the 
extent of the azimuthal angle traveled by light:
\begin{equation}
\theta_H={L_H\over a}=\int^\eta_0{{d\eta^{\prime}}\over{a(\eta^{\prime})}}.
\label{azdist}
\end{equation}
Since the coordinate volume surrounded by the comoving horizon is
\begin{equation}
{\cal V}=\int^{\theta_H}_0d\Omega_{p-1}d\theta\sin^{p-1}\theta=V_{p-1}
\int^{\theta_H}_0d\theta\sin^{p-1}\theta,
\label{cmvol}
\end{equation}
and the area of the particle horizon is
\begin{equation}
A=V_{p-1}a^{p-1}(\eta)\sin^{p-1}\theta_H,
\label{pharea}
\end{equation}
the ratio of the total entropy to the surface area is
\begin{equation}
{S_{tot}\over A}=S{{\int^{\theta_H}_0d\theta\sin^{p-1}\theta}\over{a^{p-1}
(\eta)\sin^{p-1}\theta_H}}.
\label{ratio}
\end{equation}
For example, for the $p=2$ and $p=4$ cases, the ratios are respectively 
given by
\begin{eqnarray}
\left.{S_{tot}\over A}\right|_{p=2}&=&S{{1-\cos\theta_H}\over{a(\eta)\sin
\theta_H}},
\cr
\left.{S_{tot}\over A}\right|_{p=4}&=&S{{\cos 3\theta_H-9\cos\theta_H+8}
\over{12a^3(\eta)\sin^3\theta_H}}.
\label{ratiop24}
\end{eqnarray}
We can solve Eq. (\ref{azdist}) along with the first Friedmann equation 
(\ref{frdeq1}) to express $a$ as a function of $\theta_H$ in the following 
way:
\begin{equation}
a^2(\theta_H)=\bar{\cal E}\bar{\cal P}_{\phi}\sin{\theta_H\over 2},
\label{athet2}
\end{equation}
for $p=2$, and
\begin{equation}
a^{-2{{p-1}\over{p-3}}}(\theta_H)=\bar{\cal E}\bar{\cal P}^{-{{p-1}\over
{p-3}}}_{\phi}\sin\left({\pi\over 2}-{{p-1}\over 2}\theta_H\right),
\label{athepg3}
\end{equation}
for $p>3$.  In performing integration in Eq. (\ref{azdist}), we made use 
of the fact that the closed universe on the giant graviton was initially of 
zero size ($a_0=0$) and finite size ($a_0=\bar{\cal E}^{-{{p-3}\over{2(p-1)}}}
\bar{\cal P}^{1\over 2}_{\phi}$) for $p=2$ and $p>3$, respectively.  The minus 
sign in front of the $\theta_H$ term inside of the sin-function in Eq. 
(\ref{athepg3}) was fixed by using the fact that $\dot{a}\geq 0$ all the time 
for $p>3$.  First, we consider the $p=2$ case.  Eq. (\ref{athet2}) states 
that the brane universe starts from zero size ($a=0$) when $\theta_H=0$, then 
expands until it reaches the maximum radius $a_{max}=\sqrt{\bar{\cal E}
\bar{\cal P}_{\phi}}$ when $\theta_H=\pi$.  When the brane universe or the 
giant graviton reaches its maximum size (at $\theta_H=\pi$), the area $A=2\pi 
a\sin\theta_H$ of the comoving horizon vanishes.  So, the FS holographic 
bound on $S_{tot}/A$ is violated.  Namely, the gas of D0-branes forming the 
2-sphere giant graviton does not obey the FS holographic principle.  Next, 
we consider the $p>3$ case.  Eq. (\ref{athepg3}) states that the brane 
universe starts from finite size with radius $a_{min}=\bar{\cal E}^{-{{p-3}
\over{2(p-1)}}}\bar{\cal P}^{1\over 2}_{\phi}$ when $\theta_H=0$, then expands 
indefinitely, approaching infinite size ($a\to\infty$) as $\theta_H\to
\pi/(p-1)$.  Note, the comoving distance $\theta_H$ takes values within the 
interval $0\leq\theta_H<\pi/(p-1)$, only, while the brane universe expands.  
So, the surface area $A=V_{p-1}a^{p-1}\sin^{p-1}\theta_H$ never vanishes 
for $\eta>0$, implying that the FS holographic bound $S_{tot}/A<1$ can be 
satisfied by the $(p-1)$-sphere giant graviton with $p>3$.  (Note, although 
$A$ is zero initially (at $\theta_H=0$), $S_{tot}/A$ approaches a finite 
value as $\theta_H\to 0$, as can be seen from the specific explicit 
expressions for $S_{tot}/A$ in Eq. (\ref{ratiop24}).)  This is related to 
the fact that the giant graviton has negative pressure (Cf. see also Ref. 
\cite{ram} for this point).    

There are different types of cosmological entropy bounds, depending on the 
strength of gravitation of the system.  When the self-gravity of the system 
is small compared to the total energy $E$, for which the Hubble radius 
$H^{-1}=a/\dot{a}$ is larger than the radius $a$ of the universe (i.e., 
$Ha\leq 1$), the total entropy $S_{tot}$ is bounded from above by the 
Bekenstein entropy $S_B$ \cite{bek}:
\begin{equation}
S_{tot}\leq S_B\equiv {{2\pi}\over p}Ea\ \ \ \ \ \ \ {\rm for}\ \ \ \ \ \ \ 
Ha\leq 1.
\label{beb}
\end{equation}
Here, the total energy $E$ is given by the product of the energy density 
$\varrho$ and the volume $V_{tot}=V_pa^p$ of the universe.  So,  when the 
following condition is satisfied:
\begin{equation}
a^{4{{p-1}\over{p-3}}}\leq {{(p-3)^2+16}\over{(p-3)^2}}\bar{\cal E}^{-2}
\bar{\cal P}^{2{{p-1}\over{p-3}}}_{\phi},
\label{ggbeb}
\end{equation}
the total entropy of the giant graviton cannot be bigger than the following 
Bekenstein entropy of the moving giant graviton:
\begin{equation}
S_B={{(p-1)V_p}\over{128G}}\left[(p-3)^2\bar{\cal E}^2
\bar{\cal P}^{-2{{p-1}\over{p-3}}}_{\phi}a^{{p^2-1}\over{p-3}}
-(p+1)(p-7)a^{p-1}\right].
\label{ggbe}
\end{equation}
Since the closed universe on the giant graviton is not radiation dominated, 
the above Bekenstein entropy is not constant while the brane universe 
evolves, unlike the case considered in Ref. \cite{ver1}.  For a strongly 
self-gravitating universe, satisfying $Ha\geq 1$, $S_{tot}$ is conjectured 
\cite{esl,ven,kal,bou} to be bounded from above by the Hubble entropy $S_H$:
\begin{equation}
S_{tot}\leq S_H\equiv (p-1){{HV_{tot}}\over{4G}}\ \ \ \ \ \ \ {\rm for}
\ \ \ \ \ \ \  Ha\geq 1.
\label{heb}
\end{equation}
So, when the following condition is satisfied:
\begin{equation}
a^{4{{p-1}\over{p-3}}}\geq {{(p-3)^2+16}\over{(p-3)^2}}\bar{\cal E}^{-2}
\bar{\cal P}^{2{{p-1}\over{p-3}}}_{\phi},
\label{gghb}
\end{equation}
the total entropy of the giant graviton cannot be bigger than the Hubble 
entropy:
\begin{equation}
S_H={{(p-1)|p-3|V_p}\over{16G}}\sqrt{\bar{\cal E}^2\bar{\cal P}^{-2{{p-1}
\over{p-3}}}_{\phi}a^{2{{(p-1)^2}\over{p-3}}}-a^{2(p-1)}}.
\label{gghe}
\end{equation}
When $Ha=1$ or 
\begin{equation}
a=a_{crit}=\left[{{(p-3)^2+16}\over{(p-3)^2}}\right]^{{p-3}\over{4(p-1)}}
\bar{\cal E}^{-{{p-3}\over{2(p-1)}}}\bar{\cal P}^{1\over 2}_{\phi},
\label{satvala}
\end{equation}
the Bekenstein entropy (\ref{ggbe}) and the Hubble entropy (\ref{gghe}) of 
the moving giant graviton coincide.  First, for the $p=2$ case, such critical 
value is given by $a^4_{crit}=(\bar{\cal E}\bar{\cal P}_{\phi})^2/17$.  
Note, $a=L^{3/4}_2\bar{\cal P}_{\phi}r^{1/4}$ for $p=2$.  So, as the giant 
graviton moves away from the source D4-brane starting from $r=0$, the giant 
graviton initially has strong self-gravity and its total entropy is bounded 
from above by the Hubble entropy
\begin{equation}
S_H={\pi\over{4G}}\sqrt{{{\bar{\cal E}^2\bar{\cal P}^2_{\phi}}\over a^2}-a^2},
\label{gghe2}
\end{equation}
until it reaches the critical distance $r_{crit}=L^{-3}_2\bar{\cal E}^2
\bar{\cal P}^{-2}_{\phi}/17$ (corresponding to $a_{crit}$) from the source 
brane, and after the giant graviton passes through the critical distance 
$r_{crit}$ its self-gravity becomes weak and its total entropy becomes 
bounded from above by the Bekenstein entropy
\begin{equation}
S_B={\pi\over{32G}}\left[{{\bar{\cal E}^2\bar{\cal P}^2_{\phi}}\over a^3}
+15a\right].
\label{ggbe2}
\end{equation}
Second, for the $p>3$ case, $a$ is inversely related to $r$ as $a=L^{{p+1}
\over 4}_p\bar{\cal P}^{1\over{p-1}}_{\phi}r^{-{{p-3}\over 4}}$.  So, as 
the giant graviton approaches the source D$(6-p)$-brane from its maximum 
distance $r_{max}=L^{{p-1}\over{p-3}}_p\bar{\cal E}^{2\over{p-1}}
\bar{\cal P}^{-{2\over{p-1}}}_{\phi}$ (corresponding to the minimum radius 
$a_{min}=\bar{\cal E}^{-{{p-3}\over{2(p-1)}}}\bar{\cal P}^{1\over 2}_{\phi}$ 
of the brane universe), the giant graviton initially has weak self-gravity 
with its total entropy bounded from above by the Bekenstein entropy 
(\ref{ggbe}) until it reaches the critical distance (corresponding to the 
critical cosmic scale factor Eq. (\ref{satvala}))
\begin{equation}
r_{crit}=\left[{{(p-3)^2}\over{(p-3)^2+16}}\right]^{1\over{p-1}}L^{{p+1}
\over{p-3}}_p\bar{\cal E}^{2\over{p-1}}\bar{\cal P}^{-{2\over{p-1}}}_{\phi},
\label{critdist}
\end{equation}
from the source brane, and after the giant graviton passes through $r_{crit}$ 
its self-gravity becomes strong and its total entropy becomes bounded from 
above by the Hubble entropy (\ref{gghe}).  To sum up, when the distance $r$ 
between the giant graviton and the source brane is less [greater] than the 
critical distance (\ref{critdist}), the self-gravity of the giant graviton 
is strong [weak] and its total entropy $S_{tot}$ is bounded from above by 
the Hubble entropy $S_H$ [the Bekenstein entropy $S_B$].  

Verlinde proposed \cite{ver1} a new purely holographic cosmological bound 
which remains valid throughout the cosmological evolution.  We briefly 
summarize the Verlinde's proposal in the following.  It is proposed that, 
just like a CFT in a finite volume, the total energy $E$ of the closed 
universe contains a non-extensive Casimir contribution $E_C$.  The Casimir 
energy $E_C$ is defined as the violation of the Euler identity, which follows 
from the second law of thermodynamics along with an assumption that $E$ is 
extensive, in the following way:
\begin{equation}
E_C\equiv p(E+\wp V_{tot}-TS_{tot}).
\label{casendef}
\end{equation}  
The new cosmological bound puts constraint on the sub-extensive entropy, 
called a Casimir entropy $S_C$, associated with $E_C$, defined as
\begin{equation}
S_C\equiv{{2\pi}\over p}E_Ca.
\label{defcent}
\end{equation}
Since it is usually assumed that $E_C\leq E$, we have $S_C\leq S_B$.  The 
new cosmological bound for a closed universe states that
\begin{equation}
E_C\leq E_{BH},
\label{newcb}
\end{equation}
where $E_{BH}$ is the Bekenstein-Hawking energy, interpreted as the energy 
required to form a universe-size black hole and defined as the energy $E$ 
for which the Bekenstein entropy $S_B$ and the Bekenstein-Hawking entropy 
$S_{BH}\equiv (p-1){{V_{tot}}\over{4Ga}}$ are equal, namely,
\begin{equation}
{{2\pi}\over p}E_{BH}a\equiv (p-1){{V_{tot}}\over{4Ga}}\ \ \ \ \ \ 
{\rm or} \ \ \ \ \ \ 
E_{BH}\equiv p(p-1){{V_pa^{p-2}}\over{8\pi G}}.
\label{bheng}
\end{equation}
The bound (\ref{newcb}) states that the Casimir energy is not by itself 
sufficient for forming a universe-size black hole.  The new cosmological 
bound (\ref{newcb}) is equivalent to
\begin{equation}
S_C\leq S_{BH}.
\label{ncbent}
\end{equation}
$S_C$ can be interpreted as a generalization of the central charge to $p+1$ 
dimensions, with an identification $S_C\equiv{{\pi\tilde{c}}\over 6}$, where 
$\tilde{c}$ is a $(p+1)$-dimensional analogue of the central charge $c$ in 
$1+1$ dimensions.  So, the bound (\ref{ncbent}) is interpreted as a 
holographic upper limit on the degrees of freedom of the holographic dual 
theory.  From the definition of $E_{BH}$, we have the following criterion 
for a weakly and strongly self-gravitating universe:
\begin{eqnarray}
E\leq E_{BH}\ \ \ \ \ \ \ &{\rm for}&\ \ \ \ \ \ \ Ha\leq 1
\cr
E\geq E_{BH}\ \ \ \ \ \ \ &{\rm for}&\ \ \ \ \ \ \ Ha\geq 1.
\label{ebhineq}
\end{eqnarray}
The first Friedmann equation (\ref{frd1}) can be put into the following form 
of the cosmological Cardy formula:
\begin{equation}
S_H={{2\pi a}\over p}\sqrt{E_{BH}(2E-E_{BH})},
\label{coscard}
\end{equation}
which resembles the following conjectured Cardy formula, called the 
Cardy-Verlinde formula, for the holographic dual theory:
\begin{equation}
S_{tot}={{2\pi a}\over p}\sqrt{E_C(2E-E_C)}.
\label{bdcard}
\end{equation}
Eqs. (\ref{coscard},\ref{bdcard}) imply that the new bound $E_C\leq E_{BH}$ 
is equivalent to the Hubble bound $S_{tot}\leq S_H$ when $Ha\geq 1$.  The 
Hubble bound is saturated when $E_C=E_{BH}$
\footnote{This follows from the fact that $E_C\leq E_{BH}\leq E$ for 
$Ha\geq 1$ and $S_{tot}$ in Eq. (\ref{bdcard}), as a function of $E_C$, 
takes the maximum value when $E_C=E$.}, 
for which the cosmological Cardy formula (\ref{coscard}) and the Cardy 
formula (\ref{bdcard}) for the holographic dual theory coincide.  Making use 
of the identification $S_C=\pi\tilde{c}/6$, the generalized central charge 
of the holographic dual theory is therefore inferred to be 
\begin{equation}
\tilde{c}={{12}\over p}E_{BH}a_{crit}={{3(p-1)V_pa^{p-1}_{crit}}
\over{2\pi G}}, 
\label{cch}
\end{equation}
where $E_{BH}$ is evaluated at $a=a_{crit}$, for which $Ha=1$ is satisfied.  
For the weakly gravitating case, for which $Ha\leq 1$ and therefore $E_C\leq 
E\leq E_{BH}$, $S_{tot}$ in Eq. (\ref{bdcard}) is bounded from above by 
$S_B$ (which is just the Bekenstein bound (\ref{beb})) and the bound is 
saturated when $E_C=E$.  The second Friedmann equation (\ref{frd2}) can be 
put into the form
\begin{equation}
E_{BH}=p(E+\wp V_{tot}\pm T_HS_H),\ \ \ \ \ \ \ 
T_{H}\equiv\pm{\dot{H}\over{2\pi H}},
\label{2ndfeqrel}
\end{equation}
which resembles the defining relation (\ref{casendef}) for the Casimir 
energy.  The signs $\pm$ in Eq. (\ref{2ndfeqrel}) are chosen to make $T_H$ 
positive, namely, plus [minus] signs for an accelerating [a decelerating] 
universe.  Since $E_C\leq E_{BH}$ and $S_{tot}\leq S_H$ are equivalent for 
a strongly gravitating universe, Eqs. (\ref{casendef},\ref{2ndfeqrel}) 
imply the following bound on $T$ for a decelerating universe:
\begin{equation}
T\geq T_H\ \ \ \ \ \ \ {\rm for}\ \ \ \ \ \ \ Ha\geq 1.
\label{tempbd}
\end{equation}
For an accelerating universe, we have $T\geq -T_H$, which is trivially 
satisfied by any $T$ and therefore is not a bound on $T$.  The temperature 
bound (\ref{tempbd}) is saturated when $E_C=E_{BH}$.  

The above argument leading to new cosmological bounds by Verlinde does not 
depend on a specific form of the equation of state of the perfect fluid 
matter.  So, we can apply the result of Ref. \cite{ver1} to infer 
thermodynamic properties of the moving giant graviton and its holographic 
dual theory.  When the giant graviton has strong self-gravity, i.e., $r\leq 
r_{crit}$, its total entropy $S_{tot}$ is bounded from above by  $S_H$.  
This bound is saturated when $r=r_{crit}$, at which the Hubble entropy 
(\ref{gghe}) of the giant graviton takes the form:
\begin{equation}
S_{tot}=S_H={{(p-1)V_p}\over{4G}}\left[{{(p-3)^2+16}\over{(p-3)^2}}
\right]^{{p-3}\over 4}\bar{\cal E}^{-{{p-3}\over 2}}\bar{\cal P}^{{p-1}\over 
2}_{\phi}.
\label{crithe}
\end{equation}
Since the total entropy $S_{tot}$ is constant during the cosmological 
evolution, this entropy expression is expected to hold for any $a$.
Substituting Eq. (\ref{satvala}) into Eq. (\ref{cch}), we 
obtain the following generalized central charge of the holographic dual 
theory of the moving giant graviton:
\begin{equation}
\tilde{c}={{3(p-1)V_p}\over{2\pi G}}\left[{{(p-3)^2+16}\over{(p-3)^2}}
\right]^{{p-3}\over 4}\bar{\cal E}^{-{{p-3}\over 2}}\bar{\cal 
P}^{{p-1}\over 2}_{\phi},
\label{gggcc}
\end{equation}
which is proportional to $S_{tot}$ as expected from the holographic 
principle.  So, we infer that the (bosonic) degrees of freedom in the 
holographic dual theory would scale as $\sim \bar{\cal E}^{-{{p-3}\over 2}}
\bar{\cal P}^{{p-1}\over 2}_{\phi}$.  Note, the brane universe on the moving 
giant graviton decelerates [accelerates] when $p=2$ [$p>3$].  So, for the 
$p=2$ case, the temperature $T$ of the moving giant graviton at a distance 
$r\leq r_{crit}$ from the source D4-brane is bounded from below by
\begin{equation}
T_H={{3\bar{\cal E}^2\bar{\cal P}^2_{\phi}-a^4}\over{8\pi a^3
\sqrt{\bar{\cal E}^2\bar{\cal P}^2_{\phi}-a^4}}}.
\label{ggcrttmp}
\end{equation}
This bound on $T$ is saturated when $r=r_{crit}$:
\begin{equation}
T=T_H={{25}\over{16\pi}}{{17^{1/4}}\over\sqrt{\bar{\cal E}\bar{\cal 
P}_{\phi}}}.
\label{strttmp}
\end{equation}
At that point, the giant graviton has the total entropy saturated by the 
Hubble entropy
\begin{equation}
S_{tot}=S_H={\pi\over{17^{1/4}G}}\sqrt{\bar{\cal E}\bar{\cal P}_{\phi}},
\label{gghent}
\end{equation}
and the energy density and the pressure given by
\begin{equation}
\varrho={{17^{1/2}}\over{4\pi G}}{1\over{\bar{\cal E}\bar{\cal P}_{\phi}}},
\ \ \ \ \ \ \ \ \ \ \ \ \ \ 
\wp={{17^{3/2}}\over{64\pi G}}{1\over{\bar{\cal E}\bar{\cal P}_{\phi}}}.
\label{ggedprs}
\end{equation}
So, the Casimir energy of the moving giant graviton at $r=r_{crit}$ is 
$E_C=1/G$, which coincides with $E_{BH}$ in Eq. (\ref{bheng}) with $p=2$, 
as expected.  Comparing Eq. (\ref{entexp}) with the defining 
relation (\ref{casendef}) of the Casimir energy, we see that the constant 
$S_0$ is related to the Casimir energy as $E_C=-pS_0V_pT$.  For the $p=2$ 
case, by evaluating this relation at $r=r_{crit}$, we obtain $S_0=-{2\over
{25G}}{\sqrt{\bar{\cal E}\bar{P}_{\phi}}\over{17^{1/4}}}$.  The temperature 
(\ref{tmpofd0}) of the giant graviton, valid for any $a$, therefore takes the 
form:
\begin{equation}
T={{25\cdot 17^{1/4}}\over{352\pi\sqrt{\bar{\cal E}\bar{\cal P}}}}
\left[\bar{\cal E}^2\bar{\cal P}^2_{\phi}a^{-4}+5\right].
\label{temp2}
\end{equation}
The holographic dual theory  has the generalized central charge given by 
Eq. (\ref{gggcc}) with $p=2$:
\begin{equation}
\tilde{c}={6\over{17^{1/4}G}}\sqrt{\bar{\cal E}\bar{\cal P}_{\phi}}.
\label{ggcc2}
\end{equation}
For the $p>3$ case, since the brane universe on the moving giant graviton 
inflates, the temperature $T$ on the moving giant graviton at a distance 
$r\leq r_{crit}$ from the source D$(6-p)$-brane is bounded from below by 
the negative temperature $-T_H$.  So, it appears that the temperature 
bound (\ref{tempbd}) is trivially satisfied and cannot be saturated when 
$Ha=1$.  However, we find that $T$ is not always positive in the course 
of cosmic evolution on the moving giant graviton, unlike the case of the 
conventional cosmologies.  To determine integration constant $S_0$ in Eq. 
(\ref{tmpofd0}), we obtain the expression for $\left.T\right|_{r=r_{crit}}$ 
by evaluating Eq. (\ref{casendef}) at $r=r_{crit}$, at which $E_C=E_{BH}$.  
Plugging the resulting $S_0$ back into Eq. (\ref{tmpofd0}), we obtain the 
following temperature of the giant graviton, valid for any $a$:
\begin{eqnarray}
T&=&-{{p^3-7p^2+23p-1}\over{32(p^2-8p+23)\pi}}\left[{{(p-3)^2+16}\over
{(p-3)^2}}\right]^{{3-p}\over 4}\bar{\cal E}^{{p-3}\over 2}
\bar{\cal P}^{{1-p}\over 2}_{\phi}
\cr
& &\times\left[(p-3)\bar{\cal E}^2\bar{\cal P}^{-2{{p-1}\over{p-3}}}_{\phi}
a^{{p^2-p+2}\over{p-3}}+(p-7)a^{p-2}\right].
\label{tmpg3}
\end{eqnarray}
This temperature expression indeed saturates the bound (\ref{tempbd}) 
when $r=r_{crit}$, i.e., $Ha=1$.  For $3<p<7$, $T<0$ when $a^{4{{p-1}\over
{p-3}}}>{{7-p}\over{p-3}}\bar{\cal E}^{-2}\bar{\cal P}^{2{{p-1}\over
{p-3}}}_{\phi}$.  For $p\geq 7$, $T<0$ for any $a$.  Since $a^{4{{p-1}\over
{p-3}}}_{min}=\bar{\cal E}^{-2}\bar{\cal P}^{2{{p-1}\over{p-3}}}_{\phi}$, 
actually $T<0$ always in the course of the cosmic evolution for $p\geq 5$.  

In conclusion, we studied thermodynamic properties of a moving giant graviton 
in the near-horizon background of the dilatonic D-brane.  Explicit expressions 
for thermodynamic quantities of the moving giant graviton are obtained by 
applying thermodynamics of cosmology and the recently conjectured holographic 
principles of cosmologies.  This nontrivial thermodynamics is expected from 
the fact that the giant gravitons are extended objects having excitation 
spectrum arising from vibrations about their equilibrium configuration 
\cite{djm}.

\end{document}